\documentclass[twocolumn]{aastex63}
\usepackage{natbib}
\usepackage{amsmath}

\newcommand{\logg}{\ensuremath{\log g}}

\newcommand{\feh}{\ensuremath{\protect\rm [Fe/H] }}
\newcommand{\teff}{T$_{\rm eff}$}
\usepackage{graphicx}
\usepackage{tablefootnote}
\usepackage{hyperref}
\usepackage[version=4]{mhchem}
\usepackage{txfonts}
\usepackage{comment}
\usepackage{tikz}
\usepackage{epsfig}
\usepackage[caption=false]{subfig}
\usepackage[flushleft]{threeparttable}

\newcommand\aastex{AAS\TeX}

\received{\today}
\shorttitle{\aastex\ EMP stars from DESI}
\shortauthors{Allende Prieto et al.}

\begin{document}

\title{GTC Follow-up Observations of Very Metal-Poor Star Candidates from DESI}

\author[0000-0002-0084-572X]{Carlos Allende Prieto}
\affil{Instituto de Astrof\'{\i}sica de Canarias,
             V\'{\i}a L\'actea, 38205 La Laguna, Tenerife, Spain\\}
\affil{Universidad de La Laguna, Departamento de Astrof\'{\i}sica, 
             38206 La Laguna, Tenerife, Spain \\} 
\email{callende@iac.es}

\author[0000-0001-5200-3973]{David~S. Aguado}
\affil{Instituto de Astrof\'{\i}sica de Canarias,
             V\'{\i}a L\'actea, 38205 La Laguna, Tenerife, Spain\\}
\affil{Universidad de La Laguna, Departamento de Astrof\'{\i}sica, 
             38206 La Laguna, Tenerife, Spain \\} 

\affil{Dipartimento di Fisica e Astronomia, Universitá degli Studi 
di Firenze, Via G. Sansone 1, I-50019 Sesto Fiorentino}

\author[0000-0002-0264-7356]{Jonay~I. Gonz\'alez Hern\'andez}
\affil{Instituto de Astrof\'{\i}sica de Canarias,
              V\'{\i}a L\'actea, 38205 La Laguna, Tenerife, Spain\\}
 \affiliation{Universidad de La Laguna, Departamento de Astrof\'{\i}sica, 
             38206 La Laguna, Tenerife, Spain \\}

\author[0000-0003-3767-7085]{Rafael Rebolo}
\affil{Instituto de Astrof\'{\i}sica de Canarias,
              V\'{\i}a L\'actea, 38205 La Laguna, Tenerife, Spain\\}
 \affil{Consejo Superior de Investigaciones Cient\'{\i}ficas}
 \affil{Universidad de La Laguna, Departamento de Astrof\'{\i}sica, 
             38206 La Laguna, Tenerife, Spain \\}

\author[0000-0002-5758-150X]{Joan Najita}
\affil{NSF’s NOIRLab, 950 N. Cherry Ave., Tucson, AZ 85719, USA}

\author[0000-0003-1543-5405]{Christopher J. Manser}
\affil{Imperial College London, South Kensington Campus, London SW7 2AZ, UK}

\author[0000-0002-6667-7028]{Constance Rockosi}
\affil{Department of Astronomy and Astrophysics, University of California, Santa Cruz, 1156 High Street, 
Santa Cruz, CA 95065, USA}											
\affil{University of California Observatories, 1156 High Street, Sana Cruz, CA 95065, USA}

\author{Zachary Slepian}
\affil{Department of Astronomy, University of Florida, 211 Bryant  Space Sciences Center, Gainesville, FL 32611, USA}
\affil{Lawrence Berkeley National Laboratory, 1 Cyclotron Road, 
Berkeley, CA 94720, USA}

\author{Mar Mezcua}
\affil{Institute of Space Sciences, ICE-CSIC, Campus UAB, 
Carrer de Can Magrans s/n, 08913 Bellaterra, Barcelona, Spain}
\affil{Institut d’Estudis Espacials de Catalunya (IEEC), Carrer Gran Capit\`a, E-08034 Barcelona, Spain}

\author[0000-0002-6257-2341]{Monica Valluri}
\affil{Department of Astronomy, University of Michigan, Ann Arbor, MI 48109, USA}

\author{Rana Ezzeddine}
\affil{Department of Astronomy, University of Florida, 211 Bryant Space Science Center, P.O. Box 112055, Gainesville, FL 32611-2055, USA}

\author[0000-0003-2644-135X]{Sergey E. Koposov}
\affil{Institute for Astronomy, University of Edinburgh, Royal Observatory, 
Blackford Hill, Edinburgh EH9 3HJ, UK}
\affil{Institute of Astronomy, University of Cambridge, Madingley Road, 
Cambridge CB3 0HA, UK}

\author[0000-0001-8274-158X]{Andrew P.~Cooper}
\affil{Institute of Astronomy and Department of Physics, National Tsing Hua University, 101 Kuang-Fu Rd. Sec. 2, Hsinchu 30013, Taiwan}
\affil{Center for Informatics and Computation in Astronomy, NTHU, 101 Kuang-Fu Rd. Sec. 2, Hsinchu 30013, Taiwan}
\affil{Physics Division, National Center for Theoretical Sciences, Taipei 10617, Taiwan}

\author[0000-0002-4928-4003]{Arjun Dey}
\affil{NSF’s NOIRLab, 950 N. Cherry Ave., Tucson, AZ 85719, USA}

\author[0000-0002-2761-3005]{Boris T. G\"ansicke}
\affil{Department of Physics, University of Warwick, Gibbet Hill Road, 
Coventry, CV4 7AL, UK}

\author[0000-0002-9110-6163]{Ting S.~Li}
\affil{Department of Astronomy \& Astrophysics, University of Toronto, 
Toronto, ON M5S 3H4, Canada}

\author[0000-0001-6476-0576]{Katia Cunha}
\affil{Steward Observatory, University of Arizona, Tucson, AZ 85721, USA}
\affil{Institut d'Astrophysique de Paris, CNRS and Sorbonne Université, 75014 Paris, France}
\affil{Observat\'orio Nacional MCTI, S\~ao Crist\'ov\~ao, Rio de Janeiro, Brazil}

\author[0000-0002-3983-6484]{Siwei Zou}
\affil{Department of Astronomy, Tsinghua University, 30 Shuangqing Road, 
Haidian District, Beijing, China, 100190}

\author{Jessica Nicole Aguilar}
\affil{Lawrence Berkeley National Laboratory, 1 Cyclotron Road, Berkeley, CA 94720, USA}

\author{Steven Ahlen}
\affil{Physics Dept., Boston University, 590 Commonwealth Avenue, Boston, MA02215, USA}

\author{David Brooks}
\affil{Department of Physics \& Astronomy, University College London, Gower Street, London, WC1E 6BT, UK}

\author{Todd Claybaugh}
\affil{Lawrence Berkeley National Laboratory, 1 Cyclotron Road, Berkeley, CA 94720, USA}

\author[0000-0002-5954-7903]{Shaun Cole}
\affil{Institute for Computational Cosmology, Department of Physics, Durham University, South Road, Durham DH1 3LE, UK}

\author{Sarah Eftekharzadeh}
\affil{Universities Space Research Association, NASA Ames Research Centre}

\author[0000-0003-2371-3356]{Kevin Fanning}
\affil{The Ohio State University, Columbus, 43210 OH, USA}

\author{Jaime Forero-Romero}
\affil{Departamento de F\'{\i}sica, Universidad de los Andes, Cra. 1 No. 18A/10, Edificio Ip, CP 111711, Bogot\'a, Colombia}

\author[0000-0003-3142-233X]{Satya Gontcho A Gontcho}
\affil{Lawrence Berkeley National Laboratory, 1 Cyclotron Road, Berkeley, CA 94720, USA}
									
\author{Klaus Honscheid}
\affil{Center for Cosmology and AstroParticle Physics, The Ohio State University, 191 West Woodruff Avenue, Columbus, OH 43210, USA}
\affil{Department of Physics, The Ohio State University, 191 West Woodruff Avenue, Columbus, OH 43210, USA}
\affil{The Ohio State University, Columbus, 43210 OH, USA}

\author{Pascale Jablonka}
\affil{Institute of Physics, Laboratory of astrophysics, \'Ecole Polytechnique F\'ed\'erale de Lausanne (EPFL),  Switzerland }

\author{Robert Kehoe}
\affil{Department of Physics, Southern Methodist University, 3215 Daniel Avenue, Dallas, TX 75275, USA}

\author[0000-0003-3510-7134]{Theodore Kisner}
\affil{Lawrence Berkeley National Laboratory, 1 Cyclotron Road, Berkeley, CA 94720, USA}

\author{Martin Landriau}
\affil{Lawrence Berkeley National Laboratory, 1 Cyclotron Road, Berkeley, CA 94720, USA}

\author{Axel de la Macorra}
\affil{Instituto de F\'{\i}sica, Universidad Nacional Aut\'onoma de M\'exico, Cd. de M\'exico, C.P. 04510, M\'exico}

\author{Aaron Meisner}
\affil{NSF's NOIRLab, 950 N. Cherry Ave., Tucson, AZ 85719, USA}

\author{Ram\'on Miquel}
\affil{Instituci\'{o} Catalana de Recerca i Estudis Avan\c{c}ats, Passeig de Llu\'{\i}s Companys, 23, 08010 Barcelona, Spain}
\affil{Institut de F\'{i}sica d’Altes Energies (IFAE), The Barcelona Institute of Science and Technology, Campus UAB, 08193 Bellaterra Barcelona, Spain}

\author[0000-0002-2733-4559]{John Moustakas}
\affil{Department of Physics and Astronomy, Siena College, 515 Loudon Road, 
Loudonville, NY 12211, USA}

\author{Jundan Nie}
\affil{National Astronomical Observatories, Chinese Academy of Sciences, A20 Datun Rd., Chaoyang District, Beijing, 100012, P.R. China}

\author{Claire Poppett}
\affil{Lawrence Berkeley National Laboratory, 1 Cyclotron Road, Berkeley, CA 94720, USA}
\affil{Space Sciences Laboratory, University of California, Berkeley, 7 Gauss Way, Berkeley, CA  94720, USA}
\affil{University of California, Berkeley, 110 Sproul Hall \#5800 Berkeley, CA 94720, USA}

\author{Francisco Prada}
\affil{Instituto de Astrof\'{\i}sica de Andaluci\'{\i}a}

\author[0000-0001-5589-7116]{Mehdi Rezaie}
\affil{Department of Physics, Kansas State University, 116 Cardwell Hall, Manhattan, KS 66506, USA}

\author{Graziano Rossi}
\affil{Department of Physics and Astronomy, Sejong University, Seoul, 143-747, Korea}

\author{Eusebio S\'anchez}
\affil{CIEMAT, Avenida Complutense 40, E-28040 Madrid, Spain}

\author{Michael Schubnell}
\affil{Department of Physics, University of Michigan, Ann Arbor, MI 48109, USA}										
\affil{University of Michigan, Ann Arbor, MI 48109, USA}									

\author[0000-0003-3449-8583]{Ray Sharples}
\affil{Centre for Advanced Instrumentation, Department of Physics, Durham University, South Road, Durham DH1 3LE, UK}																
\affil{Institute for Computational Cosmology, Department of Physics, Durham University, South Road, Durham DH1 3LE, UK}

\author{Malgorzata Siudek}
\affil{Institute of Space Sciences, ICE-CSIC, Campus UAB, 
Carrer de Can Magrans s/n, 08913 Bellaterra, Barcelona, Spain}

\author{Verne V. Smith}
\affil{NSF’s NOIRLab, 950 N. Cherry Ave., Tucson, AZ 85719, USA}

\author{Gregory Tarl\'e}
\affil{University of Michigan, Ann Arbor, MI 48109, USA}

\author{Fiorenzo Vincenzo}
\affil{E.A. Milne Centre for Astrophysics, University of Hull, Hull, HU6 7RX, United Kingdom}
\affil{The Ohio State University, Columbus, 43210 OH, USA}

\author{Benjamin Alan Weaver}
\affil{NSF's NOIRLab, 950 N. Cherry Ave., Tucson, AZ 85719, USA}

\author[0000-0002-4135-0977]{Zhimin Zhou}
\affil{National Astronomical Observatories, Chinese Academy of Sciences, A20 Datun Rd., Chaoyang District, Beijing, 100012, P.R. China}

\author[0000-0002-6684-3997]{Hu Zou}
\affil{National Astronomical Observatories, Chinese Academy of Sciences, A20 Datun Rd., Chaoyang District, Beijing, 100012, P.R. China}

\begin{abstract}
The observations from the Dark Energy Spectroscopic Instrument (DESI) 
 will significantly increase the numbers of known extremely metal-poor stars by
 a factor of $\sim$\,10, improving the sample statistics  to study the early
chemical evolution of the Milky Way and the nature of the first stars.  
 In this paper we report high signal-to-noise  follow-up observations of
 9 metal-poor stars identified during the DESI commissioning with 
 the Optical System for Imaging and low-Intermediate-Resolution Integrated Spectroscopy (OSIRIS) 
 instrument on the 10.4\,m Gran Telescopio Canarias (GTC).  
 The analysis of the data using  a well-vetted methodology confirms 
 the quality of the DESI spectra and the performance of the pipelines 
developed for the data 
 reduction and analysis of DESI data.
\end{abstract}


\section{Introduction} \label{sec:intro}

Only  hydrogen,  helium,  and  traces  of  lithium  nuclei  were  formed  in  
 primordial  nucleosynthesis, completed 20 minutes after the Big Bang.  
Lithium and beryllium formed later in substantial  amounts  from  spallation  
processes  between  helium,  carbon,  nitrogen,  oxygen  nuclei  and  cosmic  rays, but heavier 
elements come mainly from burning in stars or explosive nucleosynthesis.  
When the  first  stars  formed  $300-500$\,Myr  after  the  Big  Bang,  they  
were  made  out  of  pristine  gas, unpolluted by metals.

Simulations show that making stars from zero-metallicity molecular clouds 
is challenging \citep[e.g.][]{greif11, stacy14}.  The fragmentation and collapse 
of the clouds is greatly enhanced by radiative cooling from carbon and oxygen 
atoms, unavailable for the first generation of stars \citep{bromm04}.  
This was thought to lead to a top-heavy stellar mass function, missing entirely 
the low-mass stars that may survive until today \citep{clark11}.  
Searches for very low metallicity stars prior to 2010 had provided very  
few  objects  at  metallicities  lower  than  
$\rm [Fe/H]=-4$ \citep{norris13I},  i.e.  
10,000  times  less  iron  than  in the Sun, in all cases with extremely large  
carbon enhancements that favored radiative cooling.

In 2011 a star was discovered with an iron abundance $\rm [Fe/H]=-5$ but 
a solar carbon-to-iron ratio  \citep{caff11}.   This  object  demonstrated  
that  star  formation  calculations  
predicting  it  was  not  possible  to  form  low-mass  stars  
with a global metallicity $\rm [Z/H]<-4$ \citep[e.g.][]{brom03, fre07} might
be inaccurate.  
Since then a second example with very similar chemistry has been 
found \citep{sta18}, and multiple theories have been proposed  
for  the  
formation  of  low-mass  stars  from  clouds  with  no  metals  
\citep[e.g.][]{stacy16}.  A single supernova event can pollute a 
volume with a radius of 10\,pc to $\rm [Fe/H]\sim-3$ \citep{deLis16}.   
Lower  metallicities  are  possible  for  lower  gas  densities  than the  
typical  values  found  in  molecular  clouds,  
an  asymmetric  supernova  explosion,  or  mass fall-back  onto a   
leftover black hole.   Fall-back,  in  particular,  can explain  the  
extreme  abundances  of carbon in the most metal-poor stars 
\citep[e.g.][]{ume03, iwamoto05, joggerst09}.

The chemical abundances of the most metal-poor stars give us critical 
information on the 
nucleosynthetic yields for the first stars and their supernovae, as well as on
the early stages of assembly of the Galaxy and its chemical evolution. 
But
given the vast volume and limited mixing in the interstellar medium in 
those early phases, a significant spread is expected and therefore 
it is essential to build a significant sample of 
extremely metal-poor (EMP) stars, and in particular those with the lowest 
metallicities.
The majority of extremely metal-poor stars associated with the Milky Way 
are expected to be part of the 
stellar halo population  
\citep[see, e.g.][]{2017MNRAS.465.2212S, 2023MNRAS.519..483C}. 
Therefore, wide area high-galactic latitude surveys are
most efficient at enlarging our samples.

There are only $\simeq$10 stars known at $\rm [Fe/H]\leq-5$, and roughly half of them 
were identified with data from the Sloan Digital Sky Survey \citep[SDSS;][]{sdss2000}. 
Other surveys such as RAVE \citep[]{rave2017}
or LAMOST \citep[]{lamost2012}, with samples of millions of stars, 
have contributed significant numbers 
of metal-poor stars, but are not 
particularly suited for exploring the most metal-poor domain.
These 
will be enhanced by new projects such as the ongoing 
GALAH \citep[]{galahdr3, 2022arXiv221005161D} and Gaia RVS \citep[]{gaiadr3},
or the upcoming WEAVE \citep[]{weave2022} and 4MOST \citep[]{4most2022} surveys. 
In addition to these, 
the Dark Energy Spectroscopic Instrument (DESI),  pursuing a
5-year survey with a primary focus on cosmology \citep[]{2013arXiv1308.0847L}, 
is now building a
spectroscopic database that will soon exceed in size those from all
its predecessors \citep[]{2016arXiv161100036D,2016arXiv161100037D,2022arXiv220510939A}. 
While the DESI survey focuses mainly on cosmological studies, a component of
the bright time survey is dedicated to observations of Milky Way targets 
\citep{2020RNAAS...4..188A,2023ApJ...947...37C}. 
Based on the yields from the SDSS, we can expect the DESI survey by 2024 
to increase the number of known stars with $\rm[Fe/H]\leq-5$ by an order of
 magnitude.

The commissioning of DESI took place between 2019 and 2020, split into two 
phases separated by a shutdown caused by the COVID-19 pandemic. After  
 Survey Validation \citep{svpaper}, the DESI survey began in May 2021.
As of this writing over 20 million spectra of galaxies and stars have 
been gathered by DESI and the observations continue at a good pace.
A number of metal-poor stars observed during the commissioning phase 
were identified for follow-up with Gran Telescopio Canarias (GTC). 
The purpose of these observations was to check the data quality and 
the performance of the DESI data reduction and analysis pipelines with
high signal-to-noise data obtained using 
a well-known instrumental setup and data analysis procedures.
In this paper we report on these observations and compare the results
from the two instruments.

\begin{table*}[]
\centering
\begin{tabular}{lllllclllll}
\hline
\hline
Name/ DESI target ID & RA(deg) & DEC(deg) & g(mag) & Texp(s) &S/N@450\,nm& Teff (K) & logg  & [Fe/H] & [C/Fe] & RV (km s$^{-1}$)\\
\hline
\hline
J1313--0019 & 198.362018 & -0.32817 & 16.4 & 3 $\times$ 1350 & 168&5294 & 1.0  & -5.03 & 3.77 & $263.9\pm15$ \\
\hline
39627757558173121 & 180.676872 & -1.23083 & 17.5 & 3 $\times$ 1200  & 100&5857 & 5.0  & -2.89 & 0.51 & $35.8\pm15$ \\
\hline
39627787731995518 & 179.145769 & -0.08047 & 18.2 & 6 $\times$ 1200 &85& 5538 & 1.0 & -3.25 & 2.02  & $-107.0\pm15$ \\
\hline
35186036195721514 & 133.137090 & 11.39200 & 18.0 & 6 $\times$ 1000 &140&  6331 & 5.0  & -3.64 & 1.80  & $-2.7\pm15$  \\
\hline
35186077543172120 & 133.380322 & 13.22537 & 16.8 & 4 $\times$ 900 &89&  4652 & 5.0  & -3.34 & -0.04  & $305.8\pm15$  \\
\hline
35186313011398455 & 157.980376 & 23.47101 & 18.2 & 3 $\times$ 1200 & 80& 6438 & 5.0  & -2.83 & 2.40  & $76.1\pm15$  \\
\hline
35186395165230796 & 168.464724 & 27.22502 & 18.1 & 6 $\times$ 1200 & 125& 6395 & 5.0  & -2.93 & $<$1.0  & $29.7\pm15$  \\
\hline
39628465179202835 & 196.004469 & 29.29881 & 17.9 & 6 $\times$ 800 &78& 4666 & 4.2 & -3.46 & 0.23  & $-47.4\pm15$  \\
\hline
39633286363875367 & 214.592028 & 52.51449 & 18.3 & 6 $\times$ 1400 & 78&5920 & 3.8  & -2.86 & $<$1.00  & $-68.8\pm15$  \\
\hline
39633315233267986 & 216.459371 & 54.39352 & 18.2 & 6 $\times$ 1200 & 121&6351 & 5.0 & -2.75 & 0.79  & $-118.6\pm15$  \\
\hline
\end{tabular}
\caption{Stars observed for this program with OSIRIS on GTC, and their inferred atmospheric parameters (see text).}
\label{tab:osiris}
\end{table*}

\section{DESI Targets}
\label{sec:desi}

The DESI  
 focal plane is set at the primary focus of the historic 
4\,m Mayall Telescope at Kitt Peak National Observatory. The focal plane 
has a field-of-view roughly 3.2 degrees in diameter 
\citep{2022AJ....164..207D}, 
populated by 5000 robotic positioners that lead as many fibers to the
desired target positions \citep[]{2023AJ....165....9S}. 
These fibers feed 10 identical spectrographs, each with 
three cameras B, R and Z, covering between 360--580 nm, 576--762 nm, and 752--982 nm,
respectively, at a FWHM resolution of about 0.18 nm (or a resolving power 
$R\equiv \lambda/{\rm FWHM}(\lambda)$ of about 2000 on the blue edge, 
increasing to nearly 6000 at the red end). A detailed 
description of the instrument
and the experiment can be found in the papers cited above, supplemented by 
several others 
already published or in preparation 
\citep[]{corrector,2017PASP..129f4101Z,2019AJ....157..168D,spec2022,dr9,redrock2022,2023AJ....165..126R,ops}.

The DESI survey prioritizes the  mapping of 
the redshift distribution of luminous galaxies
and quasars. When observing conditions make this unachievable (e.g.  when 
the moon is bright) the focus changes to nearby 
galaxies and
stars. The DESI target selection is based on the public Legacy Surveys 
\citep[]{2019AJ....157..168D}.
Preliminary target selection details were published  in 2020
\citep[]{2020RNAAS...4..188A,2020RNAAS...4..187R,2020RNAAS...4..181Z,2020RNAAS...4..180R,
2020RNAAS...4..179Y}, and several updates are available 
\citep[]{2023ApJ...947...37C,
2022arXiv220808512H,
2023AJ....165...58Z,
2023AJ....165..126R,
2023ApJ...944..107C}.
A discussion of the data quality as assessed from 
visual inspection can be found in 
\citet[]{2022arXiv220808516L,2022arXiv220808517A}. 
The DESI Early Data Release \citep[]{edr} and the Siena
Galaxy Atlas \citep[]{sga} are now publicly available.

The DESI stellar bright-time program is known as the Milky Way Survey 
\citep[]{2023ApJ...947...37C}.
Additional stellar spectra are associated to spectrophotometric 
calibrators (F- and A-type stars, plus white dwarfs), and the Backup program,
which focuses on stars $g<16$ and is active when the survey speed degrades significantly 
due to poor observing conditions, including morning and evening twilight.
There are as well some secondary programs including stars approved by
the DESI collaboration. The commissioning data discussed in this paper were
obtained under a wide range of conditions and in some cases when the instrument was not
fully functional, but the observations discussed here 
can be considered as representative of the range in the main survey.

\begin{figure*}
\begin{center}
{\includegraphics[width=190 mm]{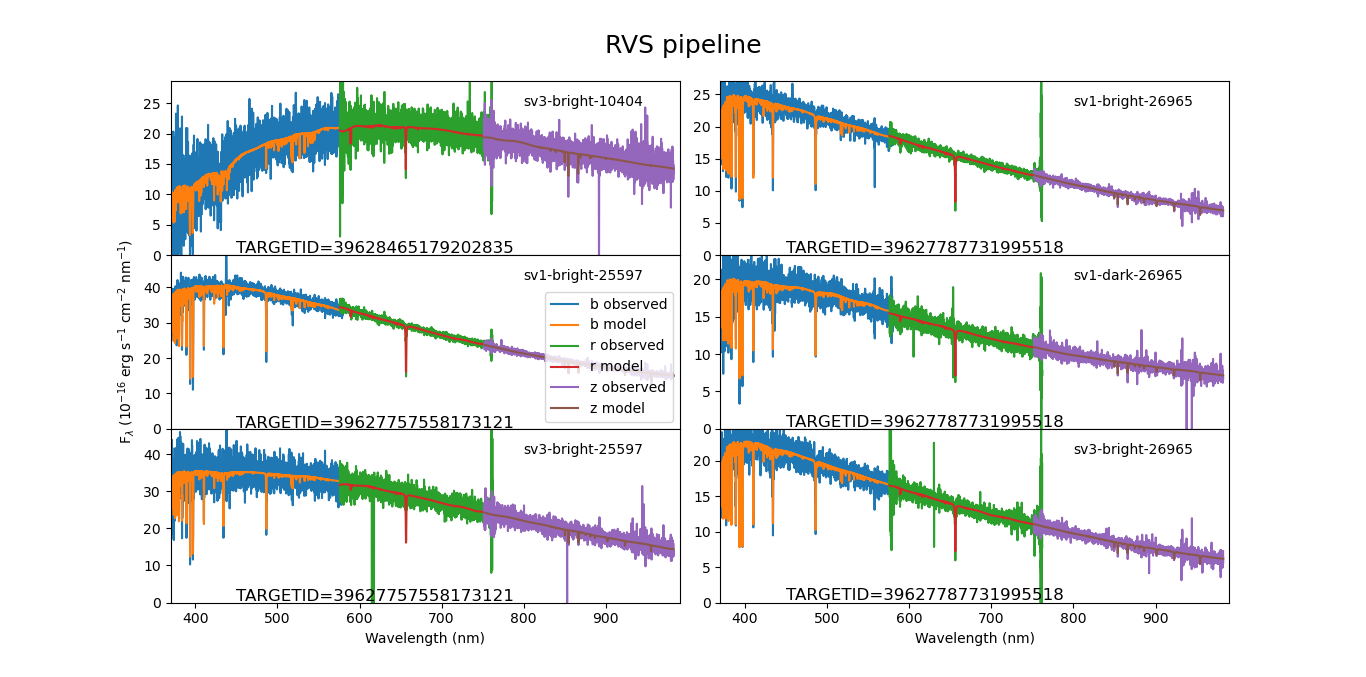}}

\end{center}
\caption{The DESI spectra of three of the candidates in our sample (blue lines) 
	and the best fits (red) from the RVS pipeline. Two of the targets 
        have been observed more than once, under different conditions or 
        target selection schemes. The numerical DESI 
        target identifications  are indicated, as well as the survey (sv1 or sv3 stand for the first and third phases 
        of the Survey Validation),  
        observing conditions or program (bright/dark sky), and healpix number 
        \citep{1999elss.conf...37G}.  All the data shown in this figure, and all the other figures
        of the paper, are available from the Zenodo repository at {\tt https://doi.org/10.5281/zenodo.8363303}.
	}
\label{fig:rvs}
\end{figure*}

Observing time on GTC was requested for 10 metal-poor stars with $g \sim 16$ 
at high 
spectral resolution using HORuS  \citep[][with a resolving power $R \sim 25,000$]{horus2021}
and 9 fainter stars with $g \sim 17-18$ at lower resolution with
OSIRIS\footnote{Optical System for Imaging and low-Intermediate-Resolution Integrated Spectroscopy} 
\citep[][$R \sim 2400$]{osiris2010} selected among the most metal-poor stars in the commissioning
data. Only the OSIRIS fraction of the requested
time was granted and these are the stars discussed in this paper. 
The sample of OSIRIS/GTC observed stars provides a good benchmark 
to test the performance of the various versions 
of the DESI stellar pipelines.

Our sample is given in Table \ref{tab:osiris} and consists of 9 stars, plus the 
ultra metal-poor star J1313$-0019$ 
\citep[]{2015A&A...579A..98A}, with [Fe/H] $\sim -5$, 
which was not observed by DESI but was nonetheless included in the
OSIRIS sample to get a solid reference for the atmospheric parameters inferred from
this instrument. 
In Fig. \ref{fig:rvs} we show in blue 6 DESI spectra of 3 of the stars studied. 
Two of these targets 
have been observed more than once under different conditions
 or target selections. The
top-left panel corresponds to DESI target ID 39628465179202835, 
 the middle and bottom left-hand 
panels correspond to target 39627757558173121,
and the three right-hand panels to 39627787731995518.
The red lines correspond to the best-fitting models from the 
DESI Milkyway Survey RVS pipeline, optimized to derive radial velocities for 
DESI data. Fig. \ref{fig:sp} shows the same spectra, 
continuum normalized as analyzed by the SP pipeline, optimized for the determination 
of chemical abundances, and the best-fitting models
from this pipeline, limited to the 370--450 nm region to facilitate visual inspection. 
These two pipelines are described in more detail in Section \ref{sec:pipeline} and 
in the overview paper by 
\citet{2023ApJ...947...37C}.

\section{OSIRIS observations}
\label{sec:observations}

The OSIRIS spectrograph mounted on the 10.4\,m GTC 
at the Roque de
los Muchachos Observatory in La Palma, Canary Islands, was used 
in longslit mode over the period 19th$-$24th January 2021. The setup  adopted grating
2500U, covering the $344-461$ nm  range, a $1.0$ arcsec slit, and $2\times2$ binning, 
leading to a resolving power of $R\sim2400$. The detector was the default device
$\rm CCD1+CCD2\_A$. To facilitate the removal of cosmic rays several exposures 
between 900 and 1400\,s were taken for each object (See Table \ref{tab:osiris}). 
As already mentioned, we also observed the hyper metal-poor star 
J1313$-$0019 \citep{alle15, frebel2015}.  In total this program, 87-MULTIPLE-2/2, 
was completed in 15 hours.
Data reduction was performed with the {\tt onedspec} package within 
 IRAF \citep{1986SPIE..627..733T, tod93}.  
Bias and flatfield correction, extraction, sky substraction, and wavelength calibration 
were performed in the standard way by using HgAr, Ne, and Xe calibration lamps. 
All the spectra were corrected for barycentric velocity calculated with {\tt rvcorrect}.

\section{Analysis of the OSIRIS-GTC observations\label{sec:analysis}}

\subsection{Radial velocities\label{sec:rvs}} 
To derive radial velocities (RVs) an OSIRIS spectrum of G64$-$12, 
a well-known metal-poor star analysed in \citet{agu17II},  
with \teff$ = 6393$~K, $\logg = 4.8$, [Fe/H]~$= -3.2$, [C/Fe]~$= +1.0$, 
was used as a template.
First, we used the IRAF task {\tt fxcor} we computed the Cross Correlation 
Function  (CCF) for each individual spectra. 
Using these first estimates of the RV of each individual spectra we 
computed an average spectrum of each star which was subsequently analysed 
with a grid of models to derive the best fit stellar parameters and 
metallicity (see Section~\ref{sec:par}). 

Due to the variety of stellar temperatures in our sample, ranging from 
about 4650~K to 6440~K, we decided to use the best-fitting FERRE synthetic spectra 
as templates to re-compute the RV of each individual spectrum~(see Fig.~\ref{fig:osiris}).
To do that we used our own IDL-based automated code, which normalizes both the 
synthetic and observed OSIRIS spectra using a running mean filter with a 
width of 30 pixels. 
The CCF was built using almost the whole OSIRIS spectral range in the 
current setup from 3760 to 4450 {\AA} with a window of 2000~km~s$^{-1}$,
in a similar manner to \citep{Arentsen2023_LAMOST_PRISTINE}.
The CCF has a similar shape as those of the strong features of the OSIRIS 
medium resolution spectra such as the H~I Balmer lines, the Ca~II H and K lines 
and the CH G-band with the adopted normalization procedure, which gives an 
oscillating pattern at the continuum of the CCF distinct from a Gaussian 
shape. To fit the whole CCF profile we employed a Gaussian model plus 
a second-order
polynomial function.
We finally derived the RV from a parabolic fit using the closest 
six points to the CCF peak. 

The statistical uncertainty of the centroid of the parabolic fit is typically 
under 1~km~s$^{-1}$, significantly below the pixel size of $\sim 0.57$
{\AA}/pixel ($\sim 42$~km~s$^{-1}$).  
We also applied a running median filter that minimizes the oscillating 
pattern and allows a better fit of the CCF using a Gaussian, although in 
this case we also adopted the parabolic fit of the CCF peak. 
Using the mean or the median normalization schemes provides differences 
under 1.5~km~s$^{-1}$ for the parabolic fits,  and under 3~km~s$^{-1}$ for the 
Gaussian fits, respectively. 
The OSIRIS spectra do not include any
telluric or sky lines useful to correct for instrument variability.
The results of the OSIRIS spectra show intra-night RV variations with 
standard deviations in the range $6.5-15$~km~s$^{-1}$, and RV variations 
from different nights with standard deviations in the range 
$4-17$~km~s$^{-1}$. Therefore, we adopted an RV uncertainty floor of 
15~km~s$^{-1}$ to the RV uncertainties, which may reflect the systematic RV 
uncertainty due to instrument flexures, pointing, guiding RV drifts, etc.
We provide the final derived RVs from
the weighted mean of the individual RVs of each target
using the normalization with a running median filter in Table~\ref{tab:osiris}.
We note that the RV of the target SDSS~J1313$-$0019 is consistent with the recent 
accurate RV of $273.984\pm 0.054$~km~s$^{-1}$ derived using the ESPRESSO 
spectrograph~\citep{aguado2022}.

\begin{figure*}
\begin{center}
{\includegraphics[width=190 mm]{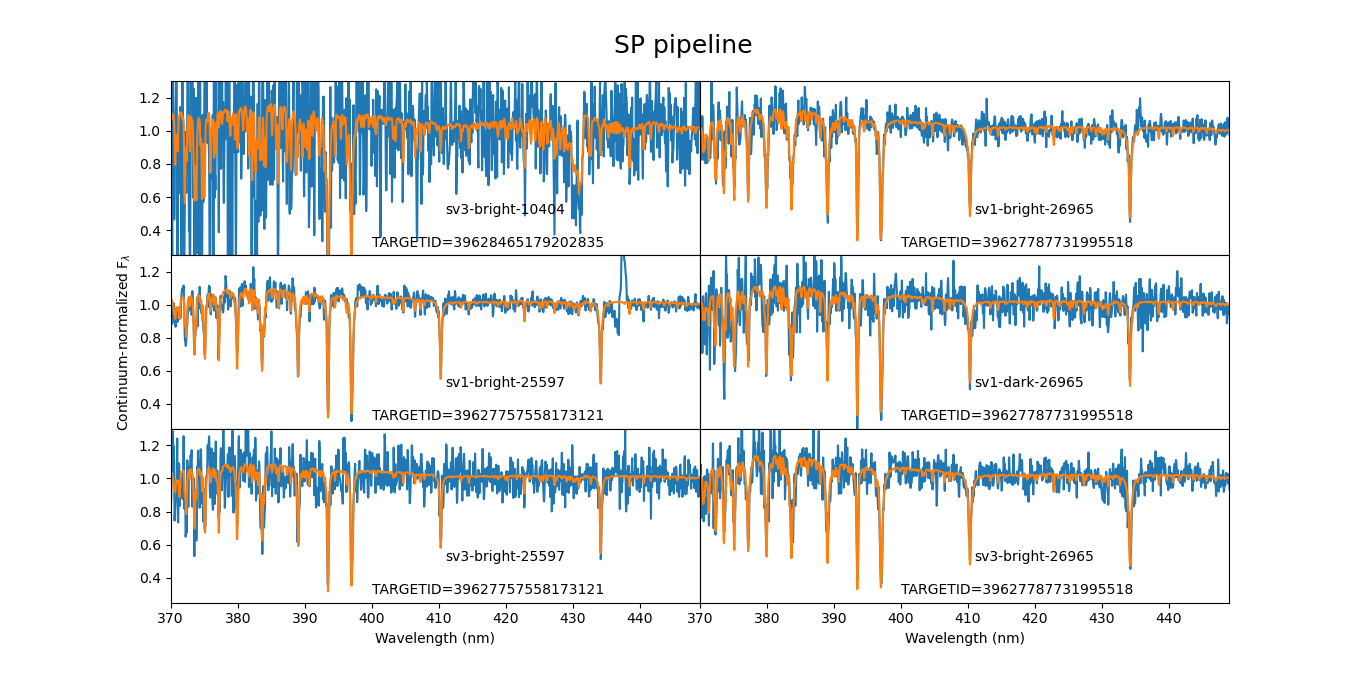}}

\end{center}
\caption{The continuum-normalized spectra of the same stars in Figure \ref{fig:rvs} (blue lines) 
	and the best fits (red) from the SP pipeline, 
	limited in this case to the 375--450 nm region.
	}
\label{fig:sp}
\end{figure*}

\subsection{Stellar parameters and metallicities\label{sec:par}} 

To derive stellar parameters, i.e., effective temperature (\teff), surface gravity (\logg), 
overall metallicity ([Fe/H]) and carbon-to-iron ratio ([C/Fe]), 
we used the FERRE 
code \citep{alle06} and followed the methodology described in \citet{agu16,agu17II, agu17I}.
We employed two similar libraries of stellar spectra spanning slightly different stellar 
parameter ranges \citep{agu17I, arentsen20} computed with the ASSET code \citep{koe08} and Kurucz model atmospheres \citep{2012AJ....144..120M}. 

In these libraries a constant value  [$\alpha$/Fe] $=+0.4$ is assumed, which  
reduces the number of free parameters to four. This approximation  is helpful, given that 
the information that can obtained at this modest resolution for very metal-poor stars is 
 mostly limited to calcium, and at the same time is valid for the 
vast majority of metal-poor stars in the Galaxy. For example, 
SMSS J160540.18-144323.1 \citep{nordlander19}, 
with [Fe/H] $\simeq -6.2$ shows [Ca/Fe] $=+0.4 \pm 0.2$, albeit larger values for Mg 
([Mg/Fe] $= 0.6 \pm 0.2$) and  Ti ([Ti/Fe] $=0.8 \pm 0.2$), and SDSS J1313-0019 
\citep{frebel2015}
with [Fe/H] $\simeq -5$ exhibits [Ca/Fe] $=0.32 \pm 0.07$. In fact, 
\citet{2023ApJ...948...38J} report calcium 
abundances for 13 star in the range $-4 <$[Fe/H]$<-2$, and they show an average [Ca/Fe] of 
$0.31 \pm 0.18$, where the quoted uncertainty corresponds to the standard 
deviation for the sample. 

The grids of synthetic spectra were smoothed to the OSIRIS resolution (R$\sim2400$). 
Then we normalized both the data and the grids, with a running mean filter of 25 pixels. 
FERRE is able to find the best fit by using the Boender-Timmer-Rinnoy-Kan (BTRK) algorithm 
\citep{boe82} interpolating between nodes of the grids to derive the best 
set of values. Both grids are basically the same but one of them goes down to metallicity 
$\rm [Fe/H]=-6$ and is suitable for ultra metal-poor stars like SDSS~J1313$-$0019.
Despite the high quality of the OSIRIS data it is not always possible to derive reliable
[C/Fe] at this resolution. Stellar parameters and carbon abundance play an important role 
here. This issue is deeply investigated in \citet{agu19b} and we use the sensitivity curves
they published and discriminate between reliable values and upper-limits. The results of 
this analysis are summarized in Table~\ref{tab:osiris}. 
The derived uncertainties ($\Delta T_{\rm eff}=\pm 120$\,K; $\Delta \logg=\pm 0.6$; 
$\Delta  \left[{\rm Fe/H}\right]=\pm 0.2$; and $\Delta  \left[{\rm C/Fe}\right]=\pm 0.3$)  
are calculated from the internal FERRE statistical errors combined with the systematics 
estimated in \citet{agu17I}. 

\begin{figure*}
\centering
{\includegraphics[width=0.75\hsize]{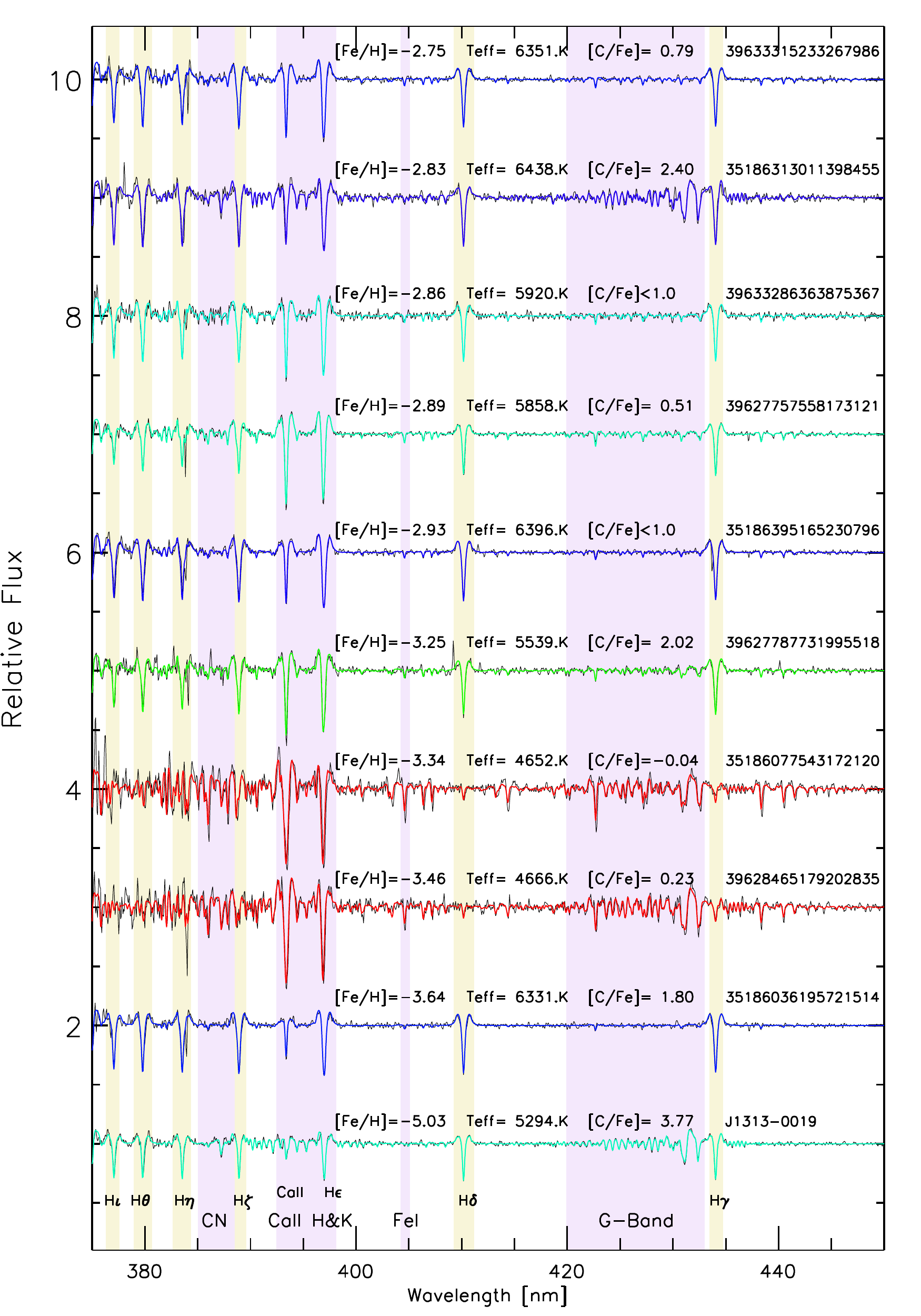}}
\caption{OSIRIS/GTC spectra (3750--4500 \AA) of our stellar sample
(black lines) and the best fits calculated with FERRE, colour-coded by \teff\,\, 
(the bluer the hotter) and sorted by decreasing \feh. The Balmer lines (yellow) 
and main metallic absorptions (purple) are highlighted. Above each spectrum 
the metallicity, effective temperature and carbon ratios are displayed. 
We also show the OSIRIS spectrum of SDSS~J1313-0019 for comparison.}
\label{fig:osiris}
\end{figure*}

In Fig.~\ref{fig:osiris} we show the OSIRIS spectra 
from our sample of EMP candidates (black lines) and the best fit derived with 
FERRE, color-coded by temperature with blue indicating warmer and red cooler.
The high quality of the OSIRIS data together with the clear CH absorptions 
in the G-band make it 
possible to derive carbon abundances and accurate metallicities 
in the majority of the cases.

For the star J1313$-$0019 we derive T$_{\rm eff}=5294\pm 120$\,K,  in good agreement 
with the values inferred by \citet{alle15, frebel2015}. The best value for gravity is 
on the edge of the grid ($\logg=1.0\pm 0.6$) while the preferred value from 
the cited original works is $\logg = 2.6$. Additionally, \citet{agu17I} obtained 
$\logg =3.6$ from a low-resolution ISIS\footnote{Intermediate-dispersion Spectrograph 
and Imaging System} spectrum. The most likely reason FERRE drifts
to a lower gravity is due to limited spectral information, given the shorter spectral 
range in the blue side for the OSIRIS data. 
Both the ISIS and BOSS\footnote{Baryon Oscillation Spectroscopic Survey} 
spectrographs provide up to 15 nm wider coverage and 
include some Balmer lines with relevant information to derive gravity. 

The impact of this deviation in \teff\ and [Fe/H] is small, and our result, 
 $\rm [Fe/H]=-5.03\pm 0.2$, is slightly lower than that of   
\citet{agu17I}, $\rm [Fe/H]=-4.7$, due the $\sim 200$\,K difference 
in temperature.
On the other hand, our inferred metallicity is consistent with those from \citet{frebel2015}, derived from \ion{Fe}{1} lines observed with the 
MIKE\footnote{Magellan Inamori Kyocera Echelle} spectrograph
on the Magellan telescope.

Finally, the amount of carbon FERRE derives,  $\rm [C/Fe]=3.77\pm0.3$, 
 is larger than the $\rm [C/Fe]=2.96$ reported by \citet{frebel2015}.
Such a difference is explained by the higher temperature (by 200\,K), 
which enhances CH dissociation and the inferred 
carbon abundance, and the lower surface gravity we derive from the 
low-resolution spectrum.
The impact of \logg~ deviations when measuring 
carbon from low-resolution spectroscopy was deeply studied in \citet{agu19b} 
and tends to increase the amount of carbon when the gravity is underestimated, 
consistent with what we find. If we force FERRE to include $\logg=2.6$ then
we recover the same [C/Fe] ratio.

\section{Results from the DESI pipelines}
\label{sec:pipeline}

All DESI data are reduced by a pipeline that performs 
wavelength calibration,
flatfielding, and spectral extraction accounting for the 2D shape of the PSF, 
flux calibration, fiber cross-talk correction, combination of
multiple exposures, and spectral classification and redshift/radial velocity
determination \citep[]{spec2022}. 

The stellar spectra are also processed by three pipelines developed by the Milky Way
Survey working group to determine stellar radial velocities, atmospheric parameters
and chemical abundances: the RVS,  SP, and WD pipelines \citep[]{2023ApJ...947...37C}. 
The RVS pipeline derives radial velocities and atmospheric parameters (\teff), 
$\log g$, [Fe/H] and [$\alpha$/Fe]) using the RVSPEC code \citep[]{rvspec}, based on Phoenix models \citep{2013A&A...553A...6H}. 
Similarly, the SP pipeline
derives atmospheric parameters and the abundances of several individual elements 
by means of the FERRE code \citep[]{2006ApJ...636..804A}. We should note that FERRE is the code adopted for 
the analysis of GTC spectra, however, the configuration chosen and actual models adopted
differ between the two analyses. The WD pipeline is focused on the classification and 
the determination of the atmospheric parameters (\teff\ and $\log g$) of white
dwarfs.


The DESI reduction pipeline is continuously being improved, 
and several internal data releases have been produced which are named after 
mountains or mountain ranges: {\it Andes, Blanc, Cascades, Denali, Everest, Fuji}. 
{\it Fuji} has been tagged for public liberation in the Early Data Release 
\citep{edr} and contains commissioning and survey validation 
observations. 
The Milky Way Survey pipelines have also been evolving with these 
internal data releases, and we present the resulting analyses for each 
release in Table \ref{tab:desi}. We have disregarded the {\it andes} and {\it cascade} 
releases from our discussion since we did not have results from both the 
SP and RVS pipelines for them.

\begin{table*}[]
\centering
\begin{tabular}{llllllllllllllll}
\hline
\hline
\multicolumn{1}{c}{} &
\multicolumn{3}{c}{OSIRIS} &  
\multicolumn{3}{c}{Blanc} & 
\multicolumn{3}{c}{Denali} & 
\multicolumn{3}{c}{Everest} & 
\multicolumn{3}{c}{Fuji}  \\
Name & Teff & logg & [Fe/H] & Teff & logg & [Fe/H] & Teff & logg & [Fe/H] & Teff & logg & [Fe/H] & Teff & logg & [Fe/H] \\
\hline
\hline
\multicolumn{16}{c}{RVS} \\
39627757558173121 & 5857 & 5.0  & -2.89 & 5935 & 4.82 & -3.00 & 5938 & 4.85 & -2.97 & 5848 & 4.70 & -3.04 & 5842 & 4.57 & -2.86 \\
39627787731995518 & 5538 & 1.0 & -3.25  & 6138 & 4.84 & -2.69 & 6167 & 4.66 & -2.91 & 6181 & 4.78 & -2.68 & 6162 & 4.80 & -2.67 \\
35186036195721514 & 6331 & 5.0  & -3.64 & 6484 & 4.44 & -4.00 & \dots & \dots & \dots & \dots & \dots & \dots & \dots & \dots & \dots \\
35186077543172120 & 4652 & 5.0  & -3.34 & 4764 & 4.63 & -2.89 & \dots & \dots & \dots & \dots & \dots & \dots & \dots & \dots & \dots \\
35186313011398455 & 6438 & 5.0  & -2.83 & 6249 & 4.38 & -2.25 & \dots & \dots & \dots & \dots & \dots & \dots & \dots & \dots & \dots \\
35186395165230796 & 6395 & 5.0  & -2.93 & 6513 & 4.85 & -3.00 & \dots & \dots & \dots & \dots & \dots & \dots & \dots & \dots & \dots \\
39628465179202835 & 4666 & 4.2 & -3.46 & 4811 & 4.06 & -3.00 & \dots & \dots & \dots & 5071 & 5.93 & -3.05 & 4812 & 4.41 & -3.03 \\
39633286363875367 & 5920 & 3.8  & -2.86 & 6281 & 4.65 & -3.02 & 6444 & 5.00 & -2.60 & 6348 & 4.78 & -2.62 & 6378 & 4.78 & -2.58 \\
39633315233267986 & 6351 & 5.0 & -2.75 & 8190 & 3.78 & -2.10 & 6563 & 4.97 & -2.46 & 6463 & 4.78 & -2.67 & 6441 & 4.72 & -2.69 \\
\hline
\multicolumn{16}{c}{SP} \\
39627757558173121 & 5857 & 5.0  & -2.89 & 5942 & 5.00 & -2.85 & 5841 & 5.00 & -2.99 & 5819 & 4.65 & -3.22 & 5822 & 4.81 & -2.99 \\
39627787731995518 & 5538 & 1.0 & -3.25  & 6000 & 4.87 & -2.75 & 6189 & 4.99 & -2.68 & 6122 & 4.79 & -2.65 & 6115 & 4.77 & -2.65 \\
35186036195721514 & 6331 & 5.0  & -3.64 & 6508 & 4.65 & -3.68 & \dots & \dots & \dots & \dots & \dots & \dots & \dots & \dots & \dots \\
35186077543172120 & 4652 & 5.0  & -3.34 & 4636 & 4.94 & -2.93 & \dots & \dots & \dots & \dots & \dots & \dots & \dots & \dots & \dots \\
35186313011398455 & 6438 & 5.0  & -2.83 & 6325 & 4.38 & -3.05 & \dots & \dots & \dots & \dots & \dots & \dots & \dots & \dots & \dots \\
35186395165230796 & 6395 & 5.0  & -2.93 & 6385 & 4.77 & -2.93 & \dots & \dots & \dots & \dots & \dots & \dots & \dots & \dots & \dots \\
39628465179202835 & 4666 & 4.2 & -3.46  & 4460 & 4.17 & -3.48 & \dots & \dots & \dots & 4743 & 5.00 & -3.11 & 4790 & 4.91 & -3.04 \\
39633286363875367 & 5920 & 3.8  & -2.86 & 5694 & 4.16 & -1.66 & 6342 & 4.89 & -2.60 & 6336 & 4.84 & -2.72 & 6360 & 4.84 & -2.56 \\
39633315233267986 & 6351 & 5.0 & -2.75  & 5513 & 3.99 & -2.45 & 6456 & 4.84 & -2.62 & 6421 & 4.81 & -2.81 & 6418 & 4.79 & -2.80 \\
\hline
\end{tabular}
\caption{Atmospheric parameters derived by the DESI Milky Way Survey RVS and SP pipelines for various data releases.  The missing data are related to commissioning observations not processed as part of internal data releases after {\it Blanc}.  
The parameters derived from the OSIRIS spectra are copied from Table \ref{tab:osiris} for reference.}
\label{tab:desi}
\end{table*}

Since the stars chosen for follow-up observations with GTC were all observed in the 
DESI commissioning, and some of the data were affected by issues that limited the
instrument performance, all of them appear in the catalogs from the early data sets 
(e.g. {\it Blanc}) but some are missing in the newer ones, which only include those 
reobserved after commissioning. 
In our analysis we have only considered coadded spectra from any given program.
Individual exposures taken within a survey/program pair (e.g. Survey Validation 1, 2 or 3; and 
the Backup program) were coadded, but data from different survey/program pairs analyzed 
independently. We further averaged multiple results obtained for any given star, 
reducing them down to a single entry per star in Table \ref{tab:desi}.

\section{Results and discussion}\label{sec:discussion}

A quick inspection of Figs. \ref{fig:rvs} and \ref{fig:sp} reveals that the DESI observations 
span a significant range in signal-to-noise ratio. This happens naturally due to the
broad magnitude range and single-valued effective exposure times 
\citep{spec2022} for the entire focal plane, but
also due to the existence of multiple programs that take stellar spectra  (dark time
observations, Milky Way Survey, Backup program, etc.). Changing observing conditions are
an additional source of spread in signal-to-noise ratios, but the use of an exposure-meter
mitigates this effect.

The follow-up data taken at GTC and illustrated in Fig. \ref{fig:osiris} 
have substantially higher signal-to-noise
ratios. 
One should also bear in mind
that the analysis of both DESI and the OSIRIS observations suffer from similar 
theoretical shortcomings when it comes to producing accurate atmospheric parameters and
abundances, such as departures from local thermodynamical equilibrium 
and hydrostatic equilibrium \citep{2009LRSP....6....2N,
2016LRSP...13....1A},
as well as uncertainties in the atomic/molecular data or other systematics affecting the construction
of the model atmospheres adopted. These issues are necessarily left out of the comparison,
with the hope and expectation that they will not prevent us from confirming that DESI data
can deliver observations free from systematic errors, at least at the same level as other
existing state-of-the-art instruments.

\subsection{Atmospheric parameters}
\label{sec:params}

Looking at the results in Table \ref{tab:desi} and comparing with the reference atmospheric
parameters included in Table \ref{tab:osiris} it is apparent that the early ({\it Blanc}) results
for some of the targets, notably target ID 39633315233267986, were fairly discrepant 
from those obtained in the more recent versions, which are better aligned with the reference
values. This is likely related to a number of improvements made in the 
data reduction pipeline after {\it Blanc}.

\begin{table*}[]
\centering
\begin{tabular}{ccccccccc}
\hline
\hline
dataset & pipeline & Nstars  & $< \Delta T_{\rm eff} >$ & $\sigma$(Teff) & $< \Delta \log g >$ & $\sigma$( $\log g$) & $< \Delta$ [Fe/H] $>$ & $\sigma$(Fe) \\
        &          &        &  K                     &       K        &       cm s$^{-2}$    &  cm s$^{-2}$ &   dex  &  dex  \\
\hline
\hline
Blanc  & RVS & 9 &  265.42 & 581.77 & -0.44 & 0.37 & 0.15 & 0.35 \\
Blanc  & SP  & 9 &  -168.23 & 284.25 & -0.39 & 0.31 & 0.17 & 0.38 \\ \hline
Denali & RVS & 4 &  209.10 & 101.65 & -0.03 & 0.26 & 0.09 & 0.18 \\
Denali & SP  & 4 &  137.98 & 92.94 & 0.02 & 0.16 & 0.10 & 0.14 \\ \hline
Everest& RVS & 5 &  170.74 & 101.30 & 0.10 & 0.53 & 0.12 & 0.17 \\
Everest& SP  & 5 &  76.54 & 104.39 & -0.07 & 0.23 & 0.03 & 0.22 \\ \hline
Fuji   & RVS & 5 &  115.26 & 102.31 & -0.23 & 0.21 & 0.17 & 0.12 \\
Fuji   & SP  & 5 &  89.52 & 102.32 & -0.07 & 0.18 & 0.13 & 0.18 \\
\hline
\end{tabular}
\caption{Statistics (mean and standard deviation) for the differences between 
the parameters derived from various data releases and pipelines (RVS or SP) with respect to those obtained 
for the same stars from the OSIRIS observations. 
Going down in the table leads to more recent data sets.}
\label{tab:stats}
\end{table*}

Table \ref{tab:stats} shows the mean and standard deviation of the differences between
the various results from the DESI pipelines and data releases,  and those obtained from the analysis of the
OSIRIS data. The DESI spectrographs
are fiber-fed and rest in a  temperature-controlled room. Furthermore, the presence
of prominent telluric lines in the wide DESI spectral range allows for a determination
of possible zero points. On the other hand, OSIRIS was mounted on the rotating
Nasmyth-B focus of the GTC, subject to variable forces and temperatures. 
(Since then the instrument
has been moved to the Cassegrain focus, but the same caveats hold). The velocities
from DESI are the reference here and the significant variations found are expected
and attributed to systematics in the OSIRIS data. Therefore, for clarify, radial velocities are
not included in the comparison in Table \ref{tab:stats}.

The atmospheric parameters from DESI and OSIRIS show significant discrepancies
in the oldest data release considered ({\it Blanc}), but then fairly good agreement in all the 
following ones. This is likely due to 
i) the removal of commissioning observations for internal data releases after  
 {\it Blanc}, and ii) the evolution of the software in the pipelines,
both in the data reduction pipeline as well as in the Milky Way Survey RVS and SP 
stellar analysis pipelines.

The most relevant statistics for our study are the most recent ones, those corresponding
to {\it Fuji} --the data released in the DESI Early Data
Release \citep{edr}. These results indicate that there is good agreement, with small zero-point 
differences and modest scatter, both at the level of 100 K for \teff,
and $<0.2$ dex for $\log g$ as well as [Fe/H]. This is true for both the RVS and SP
pipelines. 

The low scatter is smaller than the uncertainties expected for the OSIRIS observations
 (the latter are 120 K for \teff, 0.6 dex for $\log g$ and 0.2 dex for [Fe/H]), however
 the expected uncertainties include an estimate of systematic errors 
  due to the approximations 
 involved in the analysis, and some of these cancel out in the statistics given
 that they apply to both the analysis of the OSIRIS and DESI spectra.  In
 addition, all the DESI 
 targets considered in this study appear to be dwarf stars, which may hide systematic
 errors affecting giants, such as those discussed in \S \ref{sec:analysis} for 
 J1313$-$0019.

\subsection{Elemental abundances}
\label{sec:carbon}

The most recent version of the SP pipeline, which was run on the {\it Fuji} 
 data release, includes
a preliminary determination of individual abundances (C, Mg, Ca and Fe),
and therefore we have estimates of the carbon and iron abundances for the stars under study 
from DESI spectra. The analysis uses the same model spectra as for the derivation of
atmospheric parameters, but for any given element the main atmospheric parameters are frozen
and only one parameter ([$\alpha$/Fe] for $\alpha$ elements, or [Fe/H] for the rest of the
elements) is searched, computing the $\chi^2$ using only sections of the spectra
dominated by transitions involving that element: atomic transitions or those of a molecule 
that includes that element. Such a scheme has been successfully adopted  for the 
APOGEE survey \citep[]{aspcap16, maj17}.

We should stress that the iron abundances derived in this manner are 
fundamentally different from the overall metallicity (also named [Fe/H]) 
discussed in previous sections. The overall metallicity is determined 
from lines of all {\it metals}, i.e. all elements with atomic number higher
than 2. Despite that iron being the element that contributes most line absorption
for most late-type stars, other elements have a non-negligible weight when
performing these measurements. On the other hand, the iron abundance
discussed in this section is derived almost exclusively from the strength
of iron lines, to the extent that they are unblended with other features. In a sense
the iron abundances derived in this manner can be
considered a more pure measurement 
and should be more representative of the actual iron abundances in these stars.

\begin{figure}
\begin{center}
{\includegraphics[width=85 mm, angle=0]{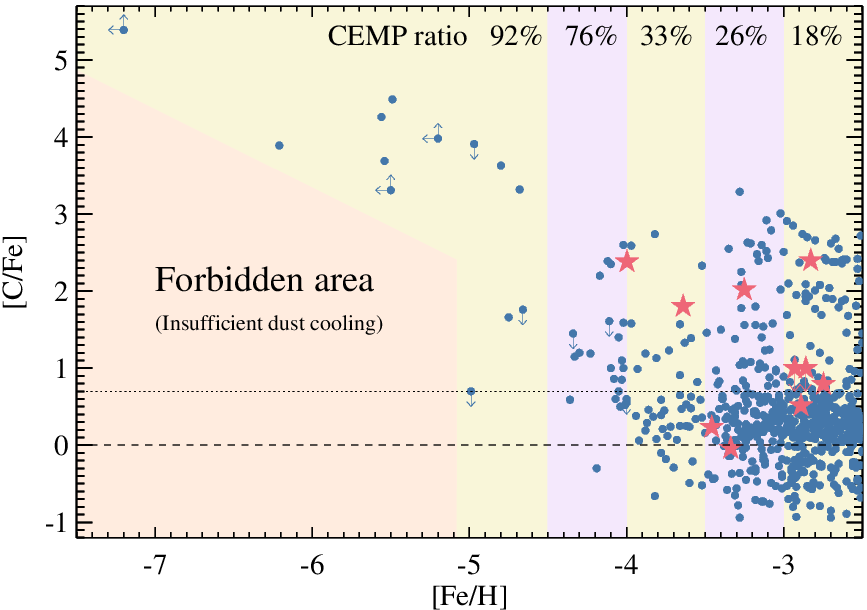}}

\end{center}
\caption{ Carbon-to-iron ratio vs. iron abundance in the ten stars from our sample 
(red stars) together with metal-poor stars with carbon abundances from the JINA database 
\citep[blue circles; ][]{jina18}. We also include the 14 known stars already known 
in the [Fe/H]$<$-4.5 regime \citet{chris04,fre05,nor07,caff11,han14,kel14,alle15,boni15,caff16,
agu18I, agu18II, agu19a, nordlander19, jon20}. The fraction of carbon-enhanced
metal-poor stars (CEMPs) in each regime are also shown. }
\label{fig:carbon}
\end{figure}

Figure \ref{fig:carbon} illustrates the  position of the targets in the [C/Fe] vs. [Fe/H] plane,
based on the OSIRIS data. Two groups are visible in this plot, one with carbon-to-iron
values at the level of about [C/Fe] $\sim 2$, and another with lower values ([C/Fe]$<1$).  
The increasing relative carbon enhancement at lower metallicity is 
considered as the signature of 
mixing and fall-back in the supernovae explosions of the first generation
of metal-free stars \citep{ume03}. The stars with lower [C/Fe] in Fig. \ref{fig:carbon} 
are not necessarily inconsistent with this picture, since the mass range 
for the progenitors and the stochastic
nature of the mixing and fall-back processes may naturally lead to a large
spread in carbon abundances \citep{boni15}. 
\citet{chiaki15} have modeled the formation 
of the first stars considering the effect of dust cooling in the protostellar
cloud, and concluded that it is not possible to form low-mass stars in the 
region higlighted in Fig. \ref{fig:carbon}. The sample of stars currently 
available is consistent with that regime, but the statistics are poor and 
DESI can play an important role to enlarge the numbers of 
extremely metal-poor stars 
with carbon determinations.  

The [C/Fe] values returned by the SP
pipeline, however, do not show an agreement nearly as good as that found for the [Fe/H] 
in Table \ref{tab:desi} for the same data release ({\it Fuji}). 
While the approximate methodology adopted is expected to break down for
large deviations from the [C/Fe]$=0$ value adopted in the construction of the model grids, 
the reasons for the discrepancies found are of a different nature. 

Early on 
during the commissioning of DESI it was found that the blue collimators of 
the DESI spectrographs have an imperfection in their coatings that induces an
optical artifact around  430 nm. The exact wavelength and the actual 
distorions induced in the spectra vary from collimator to collimator 
and somewhat as a function of time. An example of this affection can be clearly seen in 
the middle-left panel of Figure \ref{fig:sp}. This issue 
makes it difficult to use DESI spectra to infer carbon
abundances from the CH band at 430 nm, which is the main feature used
to measure the abundance of this element in medium resolution optical spectra. 
Fortunately, the DESI collimators are being replaced by others with identical
properties but without the 430 nm artifact.

The [Fe/H] values obtained with the same methodology agree quite well with the 
OSIRIS determinations with a mean difference of 0.04 and a standard deviation of 0.14 dex, 
to be compared with a mean difference of 0.13 and a standard deviation of 0.18 dex 
found for the {\it Fuji} SP DESI [Fe/H] global-parameter determinations. 

We cannot use the OSIRIS observations to evaluate the abundances 
of Mg and Ca from DESI spectra. The OSIRIS observations do not offer 
useful Mg transitions that can be 
used to derived the abundance of this element. In addition, as explained in
Section \ref{sec:par}, a constant abundance ratio [$\alpha$/Fe] is assumed
in the analysis of the OSIRIS data, which couples strongly the inferred
metallicity to the strength of the Ca II H and K lines.  

\section{Conclusion}
\label{sec:conclusion}

In this paper we report on the analysis of observations taken with the OSIRIS spectrograph
at the 10.4m GTC of a sample metal-poor stars observed by DESI. The OSIRIS data have lower spectral resolution 
and shorter wavelength coverage than those offered by DESI, but a much higher signal-to-noise
ratio and benefit from a well-vetted data treatment and analysis methodology customized
for metal-poor stars.

The targets under consideration were chosen for being part of the DESI commissioning 
observations, which in some cases included some shortcomings or issues, but some of the
targets have been since reobserved by DESI. The DESI data analysis pipelines have been
evolving continuously since the instrument started operations, and therefore multiple 
data releases, corresponding to several observing periods, programs and versions of the 
data processing software, have been considered.

We find very good agreement between the atmospheric parameters derived from 
DESI and OSIRIS data. In the most recent incarnation of the DESI data, 
released as the DESI Early Data Release, the inferred values of the stellar
\teff, $\log g$ and [Fe/H] agree with the determinations from OSIRIS data
to better than about 100 K, 0.2 dex and 0.2 dex, respectively. This applies to 
both random and systematic errors, although some of the latter may hide 
as they are shared
by the analysis of both the DESI and OSIRIS data. The results from the two DESI Milky Way Survey
pipelines, the SP and RVS branches, agree fairly well, and therefore our conclusion
applies to both. 

The preliminary values  of abundances of individual
elements provided by the DESI SP pipeline appear to be promising,
judging from the iron abundances obtained for these metal-poor stars. The 
abundances of [Fe/H] show excellent agreement with the OSIRIS values (mean 
difference of 0.04 dex and a standard deviation of 0.14 dex), 
with even higher consistency with them than the
metallicities obtained in the global simultaneous fitting of all atmospheric parameters. 
Although we also have  carbon abundances from the OSIRIS spectra, those from 
DESI exhibit significant discrepancies, which we associate with
an artifact introduced at 430 nm by the DESI collimators, which makes it 
difficult to derive reliable carbon abundances from the CH band at the
same wavelengths. The collimators are progressively being replaced, and that 
will solve this issue for future DESI observations.

\begin{acknowledgements}

Based on observations made with the Gran Telescopio Canarias (GTC), 
installed at the Spanish Observatorio del Roque de los Muchachos 
of the Instituto de Astrof\'{\i}sica de Canarias, on the island of La Palma.

CAP, JIGH and RR acknowledge financial support from the Spanish Ministry MICINN projects
AYA2017-86389-P and PID2020-117493GB-I00. Partial help was received from the program Unidad de Excelencia Mar\'{\i}a de Maeztu CEX2020-001058-M.
This research made use of computing 
time available on the high-performance computing systems at the Instituto de 
Astrof\'{\i}sica de Canarias. DA acknowledge support from the European Research Council (ERC) Starting Grant NEFERTITI H2020/808240. DA also acknowledges financial support from the Spanish Ministry of Science and Innovation (MICINN) under the 2021 Ram\'on y Cajal program MICINN RYC2021‐032609. The authors thankfully acknowledge the technical
expertise and assistance provided by the Spanish Supercomputing
Network (Red Espa\~nola de Supercomputaci\'on), as well as the computer 
resources used: the LaPalma Supercomputer and the Diva cluster, 
both located at the Instituto de Astrof\'{\i}sica de Canarias. We have made extensive use of the software from astropy\citep{2022ApJ...935..167A}. 

This material is based upon work supported by the U.S. Department of Energy (DOE), Office of Science, 
Office of High-Energy Physics, under Contract No. DE–AC02–05CH11231, and by the 
National Energy Research Scientific Computing Center, a DOE Office of Science User 
Facility under the same contract. Additional support for DESI was provided by 
the U.S. National Science Foundation (NSF), Division of Astronomical Sciences 
under Contract No. AST-0950945 to the NSF’s National Optical-Infrared Astronomy 
Research Laboratory; the Science and Technologies Facilities Council of the 
United Kingdom; the Gordon and Betty Moore Foundation; the Heising-Simons Foundation; 
the French Alternative Energies and Atomic Energy Commission (CEA); 
the National Council of Science and Technology of Mexico (CONACYT); 
the Ministry of Science and Innovation of Spain (MICINN), and by the 
DESI Member Institutions: \url{https://www.desi.lbl.gov/collaborating-institutions}. 
Any opinions, findings, and conclusions or recommendations expressed in this material 
are those of the author(s) and do not necessarily reflect the views of the 
U. S. National Science Foundation, the U. S. Department of Energy, 
or any of the listed funding agencies.

This work has made use of data from the European Space Agency (ESA) 
ission Gaia (https://www.
cosmos.esa.int/gaia), processed by the Gaia Data Processing and 
Analysis Consortium (DPAC, https:\/\/www.
cosmos.esa.int\/web\/gaia\/dpac\/consortium). Funding
for the DPAC has been provided by national institutions, 
in particular the institutions participating in the
Gaia Multilateral Agreement.

The authors are honored to be permitted to conduct scientific research on Iolkam Du’ag (Kitt Peak), 
a mountain with particular significance to the Tohono O’odham Nation.

For more information, visit desi.lbl.gov

All the data used in the figures are available from
https://doi.org/10.5281/zenodo.8363303

Software: NumPy (Harris et al. 2020), SciPy (Virtanen et al. 2020), 
Astropy (Astropy Collaboration et al.
2013, 2018), Matplotlib (Hunter 2007), PyGaia (A.
Brown; Gaia Project Scientist Support Team and the
Gaia DPAC; https://github.com/agabrown/PyGaia),
galstreams (Mateu 2017).

Facilities: ORM: GTC (OSIRIS), KPNO:Mayall (Mosaic3), Steward:Bok
(90Prime), CTIO:Blanco (DECam), WISE, Gaia.

\end{acknowledgements}

\bibliography{biblio}

\end{document}